\documentclass[12pt]{article}
\usepackage{microtype} \DisableLigatures{encoding = *, family = * }
\usepackage{graphicx,multirow}
\usepackage{amsmath}
\usepackage{amsfonts,amssymb}
\usepackage{natbib}
\bibpunct{(}{)}{;}{a}{,}{,}
\usepackage[bookmarksnumbered=true, pdfauthor={Wen-Long Jin}, breaklinks=true]{hyperref}

\newcommand{\commentout}[1]{}

\newcommand{\ba}{\begin{array}}
	\newcommand{\ea}{\end{array}}
\newcommand{\bc}{\begin{center}}
	\newcommand{\ec}{\end{center}}
\newcommand{\bdm}{\begin{displaymath}}
\newcommand{\edm}{\end{displaymath}}
\newcommand{\bds} {\begin{description}}
	\newcommand{\eds} {\end{description}}%17Apr01
\newcommand{\ben}{\begin{enumerate}}
	\newcommand{\een}{\end{enumerate}}
\newcommand{\beq}{\begin{equation}}
\newcommand{\eeq}{\end{equation}}
\newcommand{\bfg} {\begin{figure}[htbp]}
	\newcommand{\efg} {\end{figure}}%Nov 5,99
\newcommand{\bi} {\begin {itemize}}
\newcommand{\ei} {\end {itemize}}
\newcommand{\bsl} {\begin{slide}[8.8in,6.7in]}
	\newcommand{\esl} {\end{slide}}
\newcommand{\bsq}{\begin{subequations}}
	\newcommand{\esq}{\end{subequations}}
\newcommand{\bss} {\begin{slide*}[9.3in,6.7in]}
	\newcommand{\ess} {\end{slide*}}
\newcommand{\btb} {\begin {table}}
\newcommand{\etb} {\end {table}}%Nov 10,99
\newcommand{\bvb}{\begin{verbatim}}
	\newcommand{\evb}{\end{verbatim}}

\newcommand{\m}{\mbox}

\newcommand{\reff}[1] {{{Figure \ref {#1}}}}
\newcommand{\refe}[1] {{(\ref {#1})}}%Nov 5
\newcommand{\reft}[1] {{{\textbf{Table} \ref {#1}}}}
\def\pmb#1{\setbox0=\hbox{$#1$}%
	\kern-.025em\copy0\kern-\wd0
	\kern.05em\copy0\kern-\wd0
	\kern-.025em\raise.0433em\box0 }

\def\eop{{\hfill $\blacksquare$}}%17Apr01
\newtheorem{theorem}{Theorem}[section]%17Apr01
\newtheorem{definition}[theorem]{Definition}%17Apr01

\newtheorem{lemma}[theorem]{Lemma}%17Apr01
\def\dt     {{\Delta t}}

\begin {document}

\author{Wen-Long Jin\footnote{Department of Civil and Environmental Engineering, California Institute for Telecommunications and Information Technology, Institute of Transportation Studies, 4000 Anteater Instruction and Research Bldg, University of California, Irvine, CA 92697-3600. Tel: 949-824-1672. Fax: 949-824-8385. Email: wjin@uci.edu. Corresponding author}}

\title{Provably safe and human-like car-following behaviors: Part 1. Analysis of phases and dynamics in standard models} %20250428Mon11:09PDT@MBP2316

\maketitle
\begin{abstract}
Trajectory planning is essential for ensuring safe driving in the face of uncertainties related to communication, sensing, and dynamic factors such as weather, road conditions, policies, and other road users. Existing car-following models often lack rigorous safety proofs and the ability to replicate human-like driving behaviors consistently. 
This article applies multi-phase dynamical systems analysis to well-known car-following models to highlight the characteristics and limitations of existing approaches.
We begin by formulating fundamental principles for safe and human-like car-following behaviors, which include zeroth-order principles for comfort and minimum jam spacings, first-order principles for speeds and time gaps, and second-order principles for comfort acceleration/deceleration bounds as well as braking profiles. From a set of these zeroth- and first-order principles, we derive Newell's simplified car-following model. Subsequently, we analyze phases within the speed-spacing plane for the stationary lead-vehicle problem in Newell's model and its extensions, which incorporate both bounded acceleration and deceleration. 
We then analyze the performance of the Intelligent Driver Model and the Gipps model. Through this analysis, we highlight the limitations of these models with respect to some of the aforementioned principles.
Numerical simulations and empirical observations validate the theoretical insights. Finally, we discuss future research directions to further integrate safety, human-like behaviors, and vehicular automation in car-following models, which are addressed in Part 2 of this study \citep{jin2025WA20-02_Part2}, where we develop a novel multi-phase projection-based car-following model that addresses the limitations identified here.

\end{abstract}


\section{Introduction}
Drawing from the framework for behavior-based robotics \citep[][Chapter 6]{arkin1998behavior}, we can delineate driving tasks for both human-driven and autonomous vehicles into four distinct stages: communication, sensing, planning, and action. The communication and sensing stages detect, exchange, and scrutinize weather dynamics, road attributes, signage, and the dynamic presence of neighboring vehicles, bicycles, and pedestrians. Within the planning stage, the vehicle strategizes its trajectory, encompassing factors like acceleration rate and speed, navigating both longitudinal and lateral dimensions. In the action stage, an interplay of brake and engine control maneuvers the vehicle through mechanical and electrical components. 
While the action stage has been extensively studied using traditional control theory \citep[][Section 3.1]{astrom2008feedback}, advancements in communication, detection, and artificial intelligence offer the potential for significant enhancements in the communication and sensing stages, enabling offline development and testing \citep{shalev2017formal}. In contrast, the planning stage not only determines driving policies but also grapples with the complexity of environmental constraints and vehicle attributes. Thus, trajectory planning emerges as a critical component in ensuring safe vehicular operations amidst sensor uncertainties and inaccuracies.

The core component of the planning stage is a driving model, which serves as the bridge between information gathered from the communication and sensing stages and instructions relayed to the action stage. Specifically, a driving model processes real-time traffic data, road conditions, and vehicle status, and then computes and outputs trajectories, speeds, and acceleration rates to guide the action stage. It must balance a myriad of factors: ensuring safety, maintaining passenger comfort, adhering to the vehicle's physical limitations, complying with traffic laws and regulations, and adapting to ever-changing traffic conditions. Furthermore, the model must account for the vehicle's energy efficiency and consider the preferences and destinations of passengers. This multifaceted decision-making process occurs continuously, often within fractions of a second, and is crucial for both human-driven and automated vehicles. 

Automated vehicles could revolutionize transportation systems \citep{sperling2018three}, but their gradual introduction means that they will have to share the roads with human-driven vehicles and other users for the foreseeable future. Essentially, a driving model should be both safe and human-like. First, it should prevent automated vehicles from colliding with surrounding vehicles and road users when the latter's status can be measured sufficiently accurately. Second, automated vehicles should follow the same set of traffic laws and exhibit driving characteristics similar to human-driven vehicles, especially during acceleration and deceleration, so they do not surprise other road users and lead to dangerous rear-end or other collisions. Such human-like driving behaviors are also critical for automated vehicles themselves to use road infrastructure safely. For example, the change and clearance intervals for a traffic signal are designed based on vehicles' braking distances to eliminate dilemma zones; if the driving model of automated vehicles requires substantially longer braking distances, their dilemma zones may not be eliminated with current signal settings, and this could potentially endanger the safe operation of signalized intersections \citep{gazis1960problem}.

In the literature, reinforcement learning and other artificial intelligence methods have been applied to plan driving policies \citep[e.g.][]{shalev2016safe,hart2021formulation}. However, such statistical approaches cannot guarantee extremely low target fatality rates (e.g., $10^{-9}$ per hour of driving) due to the existence of many edge cases. Thus, ideally, safe and human-like driving behaviors should be mathematically provable \citep{shalev2017formal}. 
In another approach, Adaptive Cruise Control (ACC) and Advanced Driver-Assistance System (ADAS) technologies employ driving models considering various constraints like safety, comfort, vehicle characteristics, and traffic flow. Although their performance has been assessed through simulations \citep{suzuki2003effect,ntousakis2015microscopic}, mathematical safety guarantees remain elusive for such complex systems \citep{goodrich2000designing}. 
Conceptually, traditional car-following models could be implemented to control automated vehicles \citep{brackstone1999cfh,saifuzzaman2014incorporating}. These models tend to be described by transparent equations that compute a follower's speed or acceleration based on its relative position and speed to the leader. Although potentially amenable to mathematical analysis, they were generally developed to simulate traffic flow for transportation systems management and planning; whether they describe provably safe and human-like car-following behaviors has yet to be systematically studied.

In this study, we aim to develop a mathematical framework to analyze well-known car-following models as multi-phase dynamical systems and highlight the characteristics and limitations of these existing approaches. First, we systematically discuss fundamental principles for safe and human-like car-following behaviors, including zeroth-order principles for comfort and minimum jam spacings, first-order principles for speeds and time gaps, and second-order principles for comfort acceleration/deceleration bounds and braking profiles. We then analyze phases within the speed-spacing plane for the stationary lead-vehicle problem in Newell's simplified car-following model and its extensions \citep{newell2002carfollowing,daganzo2006traffic,laval2008microscopic,laval2014parsimonious,jin2018bounded}, which can be derived from a subset of the aforementioned  principles. 
Subsequently, we analyze the performance of the Intelligent Driver Model \citep{treiber2000congested} and the Gipps model \citep{gipps1981bcf}. Through this analysis, we highlight limitations of these models with respect to the defined principles.
Numerical simulations validate the theoretical insights. Finally, we discuss future research directions to further integrate safety, human-like behaviors, and vehicular automation in car-following models.

In the literature, the Wiedemann phase diagram has been used to construct psychophysical or action point models \citep{wiedemann1974simulation,fritzsche1994model,brackstone1999cfh}. These models define phases within the spacing and speed difference plane to capture perception-based behaviors of human drivers. However, a rigorous mathematical analysis framework based on such phase diagrams for car-following model evaluation has not been systematically developed or applied. In contrast, our phase diagram is based on the speed-spacing phase plane, which has been applied in empirical analyses of observed vehicle trajectories \citep{coifman2017critical} and is consistent with established speed-density and flow-density planes used in studying fundamental diagrams in steady states and other traffic scenarios \citep{jin2018bounded}.

This paper (Part 1) focuses on developing the analytical framework and evaluating existing models. In Part 2 \citep{jin2025WA20-02_Part2}, we build upon these findings to develop a novel multi-phase projection-based car-following model that addresses the limitations identified here, incorporating projected braking dynamics while maintaining parametric parsimony. Together, these two papers provide a comprehensive approach to understanding and improving car-following models for automated driving applications. The rest of the article is organized as follows. In Section 2, we define essential car-following variables and explain the principles of safe and human-like car-following behaviors. In Section 3, we analyze multiple phases in the speed-spacing plane for Newell's simplified car-following model and its extensions, especially their performance for the stationary lead-vehicle problem. In Section 4, we analyze the braking profile and the performance of the Intelligent Driver model for the stationary lead-vehicle problem. In Section 5, we analyze the steady-state solutions of the Gipps model. In 
Section 6, we conclude the study, discussing potential future research directions and the implications for developing provably safe and human-like car-following models.
\section{Definitions of variables and behavioral principles}

A list of notations is given in \reft{t:notations}.

\btb
\centering
\caption{Notation and Description}
\begin{tabular}{|c|p{10cm}|}
\hline
Notation        & Description                                       \\
\hline
$B(t)$        & Safe stopping distance                         \\
$V(\cdot)$      & Speed-density relation $v=V(k)$                                \\
$X(t)$        & Location of the follower at time $t$              \\
$X_L(t)$        & Location of the leader at time $t$              \\
\hline
$a(t)$        & Acceleration rate of the follower at time $t$     \\
$k(t)$        & Density ($k(t)=\frac{1}{z(t)}$)              \\
$q$             & Flow-rate                                         \\
$t$             & Time                                              \\
$v(t)$        & Speed of the follower at time $t$                \\
$v_L(t)$        & Speed of the leader at time $t$                \\
$v_*(t)$        & Equilibrium speed of the follower at time $t$            \\
$x$             & Distance in vehicles' traveling direction             \\
$z(t)$        & Spacing of the follower at time $t$               \\
\hline
$\Delta t$      & Time-step size                                    \\
$\alpha$        & Comfort acceleration bound of the follower            \\
$\beta$        & Comfort deceleration bound of the follower            \\
$\epsilon$      & Infinitesimal number ($=\Delta t$ in the discrete version)\\
$\kappa$        & Jam density ($=\frac{1}{\zeta}$)           \\
$\mu$          & Speed limit                                       \\
$\tau$       & Minimum time gap            \\
$\tau'$       & Reaction time for safe stopping            \\
$\zeta$      & Comfort jam spacing                \\
$\zeta'$      & Minimum jam spacing                        \\
$\zeta_c$        & Critical spacing of the follower $=\tau \mu+\zeta$ \\
\hline
\end{tabular}\label{t:notations}
\etb

\subsection{Core variables and kinematics}

\bfg\bc
\includegraphics[width=4in]{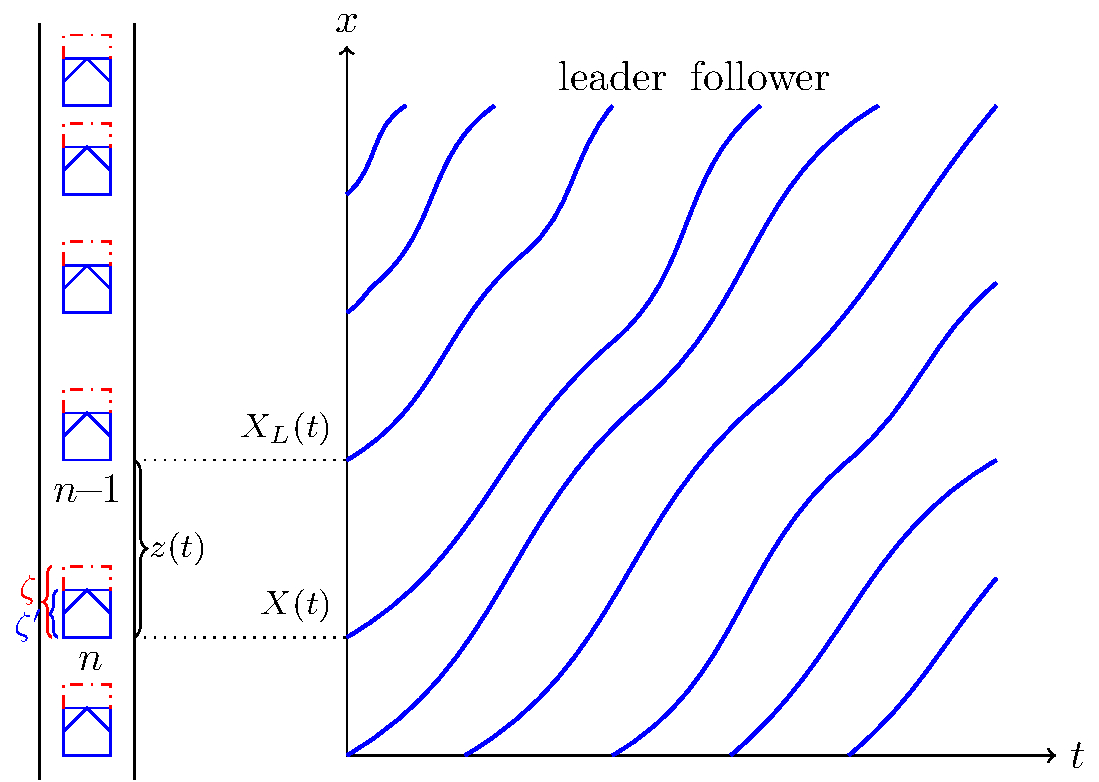}
\caption{Variables for car-following rules}\label{carfollowing-trajectories}
\ec\efg

We consider the scenario illustrated on the left side of \reff{carfollowing-trajectories}, in which a platoon of vehicles moves along a single-lane road in the positive $x$-axis direction.  For each pair of consecutive vehicles, the follower's position at time $t$ is represented as $X(t)$, measured from the rear bumper, and the leader's as $X_L(t)$. The space-time diagram on the right side of \reff{carfollowing-trajectories} displays the trajectories of these vehicles.

We define the spacing of the follower at time $t$, denoted as $z(t)$, as the difference in vehicle locations:
\begin{align}
z(t)&=X_L(t)-X(t).
\end{align}

To ensure safety, we consider two important measures: 
\begin{itemize}
\item \textbf{Minimum Jam Spacing ($\zeta'$):} The minimum required distance when stopped (approximately 5 meters for a sedan), accounting for the follower's length and a minimum safety cushion. 
\item \textbf{Comfort Jam Spacing ($\zeta$):} A typically larger spacing drivers prefer when stopped (approximately 7 meters), providing a larger comfort cushion. The term `clearance' is used to describe $z(t)-\zeta$, indicating the difference between the actual spacing and the comfort  jam spacing.
\end{itemize}

For the follower, its speed and acceleration rate are denoted by $v(t)$ and $a(t)$, respectively. If we denote an infinitesimal number by $\epsilon$, equivalent to the time-step size $\Delta t$ in the discrete version, then we can express the following relationships using the symplectic discretization scheme \citep{jin2016equivalence,jin2019nonstandard}:\footnote{This symplectic discretization scheme differs from the standard explicit Euler or ballistic update methods. It was chosen because it preserves the collision-free property in discrete implementations, while explicit Euler schemes can lead to collisions. In \citet{jin2019nonstandard}, Theorem 4.5 provides a rigorous proof showing that explicit schemes can result in collisions when a vehicle approaches a stopped leader, even if no collisions exist at previous time steps. The symplectic method avoids this critical safety issue, making it ``the only physically meaningful method'' (p.1351) for discretizing car-following models that guarantee collision avoidance.}
\bsq\label{def:symplectic}
\begin{align}
X(t+\epsilon)&=X(t)+\epsilon v(t+\epsilon), \label{def:v2X}\\
v(t+\epsilon)&=v(t)+\epsilon a(t).\label{def:a2v}
\end{align}
\esq

In a trajectory planning problem, our objective is to determine the position of the follower at the next time-step, $t+\epsilon$, denoted as $X(t+\epsilon)$, starting from the current time-step's position $X(t)$. According to \refe{def:v2X}, this task is equivalent to planning $v(t+\epsilon)$. Alternatively, we can approach it by planning the acceleration rate $a(t)$ given the current speed $v(t)$ using \refe{def:a2v}.

Additionally, the follower's time gap, denoted as $\tau(t)$, is defined as: 
\begin{align}
\tau(t)&=\frac{z(t)-\zeta}{v(t+\epsilon)}, \label{def-timegap}
\end{align}
which approaches $\frac{z(t)-\zeta}{v(t)}$ in the continuous version with $\epsilon\to0$ and is the time for the vehicle to cover the clearance with the planned speed at the next time-step.

\subsection{Behavioral principles}\label{subsection:principles}

Car-following behavior is governed by principles ensuring safety, respecting physical limits, and reflecting drivers' behaviors. These principles can be categorized by the order of derivatives of vehicle trajectories to which they apply. We now present these principles organized by both functional purpose and derivative order.

\subsubsection{Collision-free principles (Zeroth-order)}

 A meaningful car-following model should prioritize collision-free behavior. Two definitions of collision-free exist:
  \begin{itemize}
  \item \textbf{Comfort jam spacing principle:} The spacing should not be smaller than the comfort jam spacing:
    \begin{align}
    z(t) \geq \zeta. \label{CSS-principle}
    \end{align}
  \item \textbf{Minimum jam spacing principle:} In scenarios like vehicles stopping at signalized intersections with very low speeds or when they change lanes, the spacing must not be smaller than the minimum jam spacing:
    \begin{align}
    z(t) \geq \zeta'. \label{MSS-principle}
    \end{align}
  \end{itemize}
These principles relate to vehicle positions and spacings, which are zeroth-order variables in terms of vehicle trajectories.  Note that the Minimum Jam Spacing is smaller than the Comfort Jam Spacing; i.e., $\zeta' < \zeta$.

\subsubsection{Operational constraints (First-order)}

These principles relate to vehicle speeds, which are first-order derivatives of vehicle positions.

\begin{itemize}
\item \textbf{Forward traveling principle:} Assuming that all vehicles travel forward, we have:
  \begin{align} 
  v(t) \geq 0. \label{FT-principle}
  \end{align}

\item \textbf{Speed limit principle:} We assume that the planned speed does not exceed the speed limit, $\mu$: 
  \begin{align}
  v(t+\epsilon) \leq \mu. \label{SL-principle}  
  \end{align}

\item \textbf{Minimum time gap principle:} We assume the existence of a minimum time gap denoted as $\tau$, which can be expressed as $\tau(t) \geq \tau$; this condition, combined with \refe{def-timegap}, can be equivalently represented as:
  \begin{align}
  v(t+\epsilon) \leq \frac{z(t)-\zeta}{\tau}.\label{MTG-principle}
  \end{align}
  This principle connects first-order variables (speed) with zeroth-order variables (spacing) and ensures a safe following distance during movement.
\end{itemize}

\subsubsection{Acceleration constraints (Second-order)}

These principles relate to vehicle accelerations, which are second-order derivatives of vehicle positions.

\begin{itemize}
\item \textbf{Bounded control principle:} Vehicle dynamics and driver comfort are subject to bounded acceleration and deceleration:
\bsq
  \begin{align}
  -\beta \leq a(t) \leq \alpha (1- \frac{v(t)}{\mu}), \label{BC-principle}
  \end{align}
  where $\alpha$ and $\beta$ represent the acceleration and deceleration bounds. Here we adopt the TWOPAS model for bounded acceleration \citep{allen2000capability}; \citep{gipps1981bcf} proposed a slightly different model.
  
  From \refe{def:a2v}, this principle can be reformulated as:
  \begin{align}
  v(t)-\epsilon \beta \leq v(t+\epsilon) \leq v(t)+\epsilon \alpha (1- \frac{v(t)}{\mu}). \label{BC-principle-v}
  \end{align}
  \esq

\item \textbf{Safe stopping distance principle:} When the follower stops, the safe stopping distance is given by:
  \begin{align}
  B(t)&= v(t) \tau' +\frac{v^2(t)}{2\beta}, \label{SSD-principle}
  \end{align}
  where $\tau'$ is the reaction time. This safe stopping distance principle was used to derive various car-following models \citep{herrey1945principles,gipps1981bcf} and to determine change and clearance intervals for traffic signals \citep{gazis1960problem}. Note that $\tau'$ is generally smaller than $\tau$, as the former is the reaction time under braking conditions, but the latter can be considered the reaction time when vehicles drive at relatively constant speeds.
\end{itemize}

\subsubsection{Driver objectives and macroscopic properties}

\begin{itemize}
\item \textbf{Maximum speed principle:} A vehicle should aim to maximize its planned speed:
\bsq
  \begin{align}  
  \max v(t+\epsilon). \label{MS-principle}
  \end{align}
  This principle often serves as the objective function in optimization-based formulations of car-following models. From \refe{def:a2v}, this is equivalent to maximizing the acceleration rate:
  \begin{align}  
  \max a(t). \label{MA-principle}
  \end{align}  
  \esq

\item \textbf{Fundamental diagram principle:} In a steady state, when all vehicles share the same characteristics and the acceleration rate is zero, a well-defined fundamental diagram should exist:
  \begin{align}
  v=V(k), \label{FD-principle}
  \end{align}
  where $k$ represents density with $k(t)=\frac{1}{z(t)}$. This principle describes the macroscopic property of traffic flow at equilibrium.
\end{itemize}

These principles involve the following seven parameters: $\zeta$, $\zeta'$, $\tau$, $\tau'$, $\mu$, $\alpha$, and $\beta$. Their typical values except that of $\tau'$ were provided in Table 1 of \citep{treiber2000congested} and defined by ISO (International Organization for Standardization) 15622 \citep{hiraoka2005modeling}: $\zeta=7$ m, $\zeta'=5$ m, $\tau=1.6$ s, $\mu\approx 30$ m/s, $\alpha=0.73$ m/s$^2$, and $\beta=1.67$ m/s$^2$.\footnote{ISO 15622 also defines bounds for jerks, but incorporating them into our model is beyond the scope of this study.} The reaction time under braking conditions $\tau'=1$ s \citep{gazis1960problem,gipps1981bcf}. It should be noted that while we use these values for consistency with established literature, more recent trajectory data suggests higher typical acceleration values for passenger cars, ranging between 1.2-2.0 m/s$^2$ \citep{bokare2017acceleration,haas2004use}. Additionally, it is important to distinguish between the comfortable deceleration value ($\beta \approx 1.67$ m/s$^2$) used here for normal driving conditions and the emergency deceleration value ($\beta' \approx 9$ m/s$^2$ on dry roads) that would be applicable in emergency situations \citep{wang2005normal}.

Note that existing car-following models can be framed as solutions to the optimization problem, \refe{MS-principle}, subject to some or all of the aforementioned principles as well as other assumptions. 
Also note that these models implicitly assume an anisotropic traffic system, where each vehicle's primary responsibility is to avoid collisions with its immediate leader. Thus, the leader's locations at different times also serve as constraints. 
In addition, the models generally assume that initial conditions satisfy the set of principles chosen.

\section{Analysis of Newell's simplified car-following model and its extensions}
In this section, we analyze Newell's simplified car-following model and its extensions with respect to the principles outlined in the preceding section.

\subsection{Newell's simplified car-following model}

By solving   \refe{MS-principle} subject to the first-order principles in \refe{SL-principle} and \refe{MTG-principle}, we derive the following car-following model:\footnote{Without the speed limit, \refe{Newell-model} is analogous to Equation 2.3 in \citep{pipes1953operational}, although the latter was not employed to depict vehicle dynamics. In \citep{jin2016equivalence}, it was demonstrated that \refe{Newell-model} with $\epsilon=\dt$ is equivalent to the time-discrete car-following model of the Lighthill-Whitham-Richards (LWR) model with the triangular fundamental diagram \citep{lighthill1955lwr,richards1956lwr}.}
\bsq \label{Newell-model}
\begin{align}
v(t+\epsilon)&=v_*(t), 
\end{align}
where the equilibrium speed of the follower, $v_*(t)$, is defined by
\begin{align}
v_*(t)\equiv  \min\{\mu, \frac{z(t)-\zeta}{\tau}\}. \label{def:equilibriumspeed}
\end{align}
\esq
 Further, utilizing \refe{def:v2X} we can calculate the vehicle's position at the next time-step as
\begin{align}
X(t+\epsilon)&=X(t)+\epsilon v_*(t), \label{Newell-X}
\end{align}
which leads to Newell's simplified car-following model with $\epsilon=\tau$ \citep{newell2002carfollowing,daganzo2006traffic}: $X(t+\tau)=\min\{X(t)+\mu\tau, X_L(t)-\zeta\}$. 

Newell's simplified car-following model complies with all the zeroth and first order principles outlined in the preceding section, as established in the following lemmas. 

\begin{lemma} \label{lemma:Newell-CSS}
Given initial conditions satisfying the comfort jam spacing and forward traveling principles for all vehicles, i.e., $z(0)\geq \zeta$ and $v(0)\geq 0$, Newell's simplified car-following model with $\epsilon\leq \tau$ adheres to 
the comfort jam spacing and forward traveling principles at any later time $t$. 
\end{lemma}

{\em Proof}. Assuming the comfort jam spacing and forward traveling principles hold at $t$ for all vehicles; i.e., $v(t)\geq0$ and $z(t)\geq \zeta$, or equivalently, 
\begin{align*}
X(t)&\leq X_L(t)-\zeta,
\end{align*}
from \refe{Newell-model}, it is clear that $v(t+\epsilon)\geq 0$. That is, the forward traveling principle remains satisfied at $t+\epsilon$.

We further derive the following sequence of inequalities:
\begin{enumerate}
\item From \refe{Newell-X}, we have
\begin{align*}
X(t+\epsilon)&\leq X(t)+\frac{\epsilon}{\tau} (z(t)-\zeta)\\
&=X(t)(1-\frac{\epsilon}{\tau})+\frac{\epsilon}{\tau} (X_L(t)-\zeta).
\end{align*}
\item 
For $\epsilon\leq \tau$,  we conclude from the two inequalities above
\begin{align*}
X(t+\epsilon)&\leq X_L(t)-\zeta.
\end{align*}
\item Since the leader follows the forward traveling principle at $t+\epsilon$, then 
\begin{align*}
X_L(t)\leq X_L(t+\epsilon)=X_L(t)+\epsilon v_L(t+\epsilon).
\end{align*}
The two inequalities above further lead to
\begin{align*}
X(t+\epsilon)&\leq X_L(t+\epsilon)-\zeta,
\end{align*}
and, consequently, $z(t+\epsilon)\geq \zeta$. That is, the forward traveling principle also holds at $t+\epsilon$.
\end{enumerate}
Moreover, given initial conditions that fulfill the comfort jam spacing and forward traveling principles,  as $\epsilon$ is infinitesimal, by the method of induction, we conclude that 
the comfort jam spacing and forward traveling principles hold at all times.
\eop

For Lemma \ref{lemma:Newell-FD}, in \refe{Newell-model}, assuming uniform vehicle characteristics and steady states with $v(t+\epsilon)=v(t)=v$ and $z(t)=z=\frac 1k$, we can readily derive the triangular fundamental diagram in \refe{triangular-FD}.

\begin{lemma} \label{lemma:Newell-FD}
In scenarios where all vehicles share the same characteristics with $\mu$, $\tau$ and $\zeta$, Newell's simplified car-following model also complies with the fundamental diagram principle during steady states, exhibiting the following speed-density relation:
\bsq \label{triangular-FD}
\begin{align}
v&=\min\{\mu, \frac1 \tau (\frac 1k-\frac 1\kappa)\}, \label{triangular-vk}
\end{align}
where the jam density $\kappa=\frac 1\zeta$. This results in a triangular fundamental diagram:
\begin{align}
q&=k v=\{ \mu k, \frac 1\tau (1-\frac k\kappa) \}. 
\end{align}
\esq
\end{lemma}

\subsection{Extension with bounded acceleration}
From \refe{Newell-model} and \refe{def:a2v}, we have the acceleration rate in Newell's simplified car-following model as
\begin{align}
a(t)&=\frac{v_*(t)-v(t)}\epsilon, \label{Newell-a}
\end{align}
which is the nonstandard second-order formulation of the LWR model with the triangular fundamental diagram \citep{jin2019nonstandard}.
It's evident that, when the speed $v(t)$ falls below the equilibrium speed, $v_*(t)$, the acceleration rate can become quite significant, especially for small values of  $\epsilon$. To manage this, we introduce an acceleration bound in \refe{BC-principle}:
\bsq\label{BA-Newell}
\begin{align}
a(t)&=\min\{\alpha (1- \frac{v(t)}{\mu}), \frac{v_*(t)-v(t)}\epsilon\}. \label{BA-Newell-a}
\end{align}
This addition leads to the following expressions:
\begin{align}
v(t+\epsilon)&=\min\{v(t)+\epsilon \alpha (1- \frac{v(t)}{\mu}), \mu, \frac{z(t)-\zeta}{\tau}\}, \label{BA-Newell-v}\\
X(t+\epsilon)&=X(t)+ \epsilon \min\{v(t)+\epsilon \alpha (1- \frac{v(t)}{\mu}), \mu, \frac{z(t)-\zeta}{\tau}\}. \label{BA-Newell-x}
\end{align}
\esq
We refer to this modified model as the BA-Newell model \citep{jin2018bounded}.

The BA-Newell model is a direct consequence of the maximum speed, minimum time gap, speed limit, and bounded acceleration principles. Moreover, as demonstrated in the following lemmas, it consistently upholds all other zeroth- and first-order principles. Due to the similarity to the proofs presented in Lemmas \ref{lemma:Newell-CSS} and \ref{lemma:Newell-FD}, we omit the proofs of Lemmas \ref{lemma:BA-Newell-CSS} and \ref{lemma:BA-Newell-FD}.

\begin{lemma} \label{lemma:BA-Newell-CSS}
Given initial conditions that satisfy the comfort jam spacing and forward traveling principles for all vehicles, the BA-Newell model with $\epsilon\leq \tau$ remains aligned with the 
comfort jam spacing and forward traveling principles at subsequent times $t$. 
\end{lemma}

\begin{lemma} \label{lemma:BA-Newell-FD}
Under the conditions where all vehicles share identical characteristics denoted by $\mu$, $\tau$, and $\zeta$, the model also aligns with the fundamental diagram principle during steady states, showcasing the same triangular fundamental diagram as defined in \refe{triangular-FD}.
\end{lemma}

In the BA-Newell model, the traffic state of the follower can be categorized into four distinct phases: the bounded acceleration phase when  $v(t)\leq v_*(t) -\epsilon \alpha (1- \frac{v(t)}{\mu})$ and $a(t)=\alpha(1- \frac{v(t)}{\mu})$; the equilibrium cruising phase when  $v(t)= v_*(t) $ and $a(t)=0$; the equilibrium acceleration phase when $ v_*(t) -\epsilon \alpha(1- \frac{v(t)}{\mu})< v(t)<v_*(t)$ and $0<a(t)<\alpha(1- \frac{v(t)}{\mu})$; and the equilibrium deceleration phase when $v(t) >  v_*(t)$ and $a(t)<0$. The four phases are visually represented by the green solid line (bounded acceleration), blue dot (equilibrium cruising), cyan dash-dotted line (equilibrium acceleration), and brown dashed line (equilibrium deceleration) in the $(v(t),a(t))$ phase plane, as depicted in \reff{BA-Newell-4phases-new}(a). In \reff{BA-Newell-4phases-new}(b), the four phases are also illustrated by the correspondingly colored regions or curve in the $(v(t),z(t))$ phase plane, where the four phases are delineated by the following regions (here the critical spacing is defined by $\zeta_c\equiv \mu\tau+\zeta$):
\begin{itemize}

\item Bounded acceleration: $\{(v(t),z(t))| 
    v(t)\leq \mu-\epsilon \alpha(1- \frac{v(t)}{\mu}),
    z(t)\geq \tau v(t)+\zeta+\epsilon \tau \alpha(1- \frac{v(t)}{\mu})\}$.

\item Equilibrium cruising: $\{(v(t),z(t))|
    v(t)= \mu,
    z(t)\geq \zeta_c; \m{ or }
    v(t)\leq \mu,
    z(t)= \tau v(t)+\zeta\}$.
  
\item Equilibrium acceleration: $\{(v(t),z(t))| 
    v(t) < \mu,
    z(t) > \tau v(t)+\zeta;\m{ and }
    \{v(t)> \mu-\epsilon \alpha(1- \frac{v(t)}{\mu}), \m{ or }
    z(t)\geq \tau v(t)+\zeta+\epsilon \tau \alpha(1- \frac{v(t)}{\mu})\}\}$.
\item Equilibrium deceleration: $\{(v(t),z(t))| 
    v(t) > \mu,
    z(t) \geq \zeta_c;
    \m{ or }
    z(t)< \tau v(t)+\zeta,
    z(t)\leq \zeta_c\}$.

\end{itemize}
In \reff{BA-Newell-4phases-new}(b), initial values within the grey region are not permissible as they lead to negative speeds. Thus, the initial conditions have to satisfy  \refe{CSS-principle}; otherwise, the forward traveling principle in \refe{FT-principle} is violated. Nevertheless, vehicles may find themselves in these scenarios when coming to a halt at an intersection or switching lanes. 

\bfg\bc \includegraphics[width=5.5in]{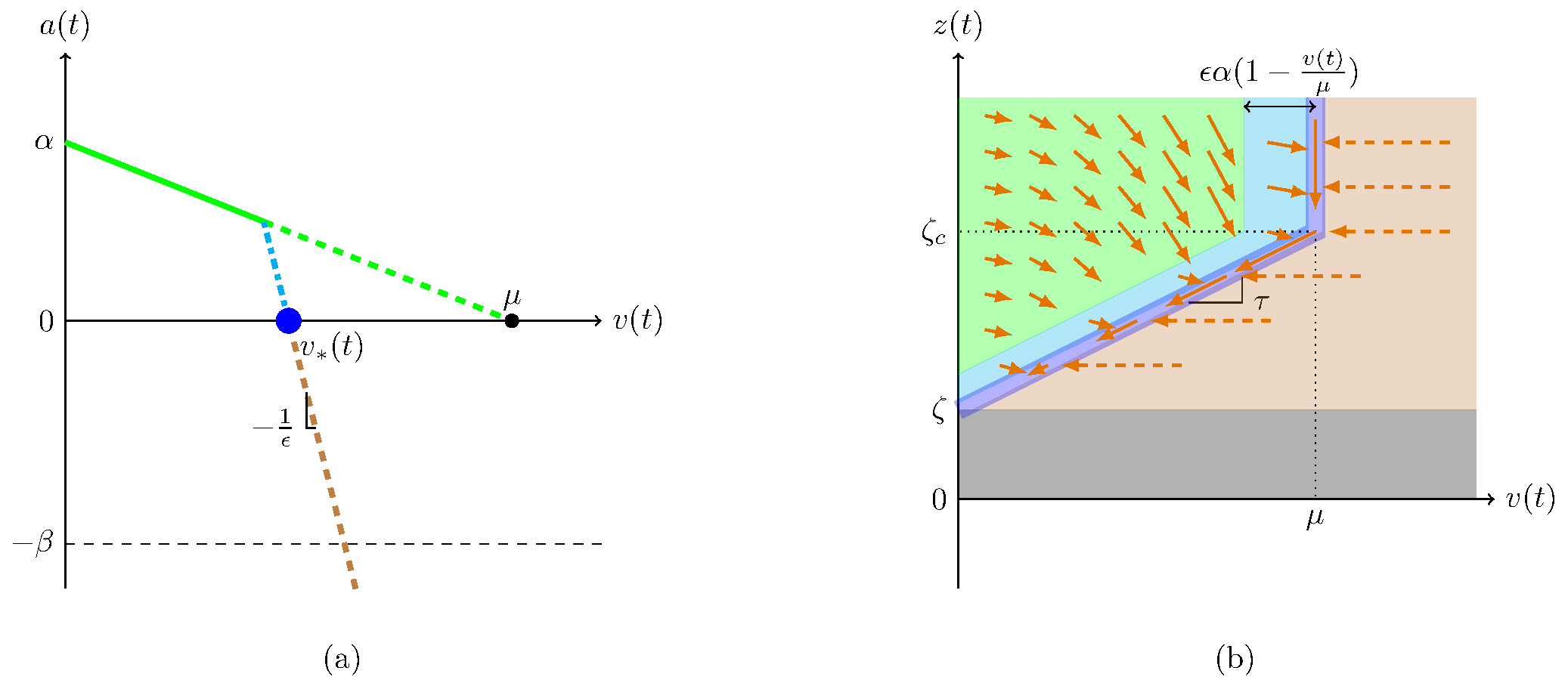} \caption{Four phases in the BA-Newell model}\label{BA-Newell-4phases-new} \ec\efg

\begin{definition}[Stationary lead-vehicle problem]
In the stationary lead-vehicle problem, the leader remains stationary with $v_L(t)=0$ at all times.
\end{definition}

In \reff{BA-Newell-4phases-new}(b), we illustrate the vector field in the $(v(t), z(t))$ phase plane for the stationary lead-vehicle problem. The arrows in this plot visually represent the changes in these variables and correspond to the vectors of $(\frac{d}{dt}v(t),\frac{d}{dt}z(t))$. In the stationary lead-vehicle problem, $\frac{d}{dt} z(t)=-v(t)$, which is independent of the leader's dynamics. Notably, within the equilibrium deceleration phase, the deceleration rate $-a(t)$ can attain significantly high initial values, indicated by the dashed arrows in \reff{BA-Newell-4phases-new}(b), so that $v(t)$ approaches values near $v_*(t)$ during one time step.  This behavior is illustrated in \reff{fig:ba_newell_slvp_ec}, where $\zeta=7$ m,  $\tau=1.6$ s, $\mu= 30$ m/s, $\alpha=0.73$ m/s$^2$, and $\epsilon=\Delta t=0.001$ s. Therefore, the BA-Newell model converges to the equilibrium states (equilibrium acceleration, equilibrium cruising, and equilibrium deceleration), in which $v(t)=v_*(t)$, regardless of initial conditions. However, there is no  bound on the deceleration rate, as seen in \reff{fig:ba_newell_slvp_ec}(a), where the deceleration rate can be as large as $18.75$ m/s$^2$.

\bfg
\centering
  \includegraphics[width=6in]{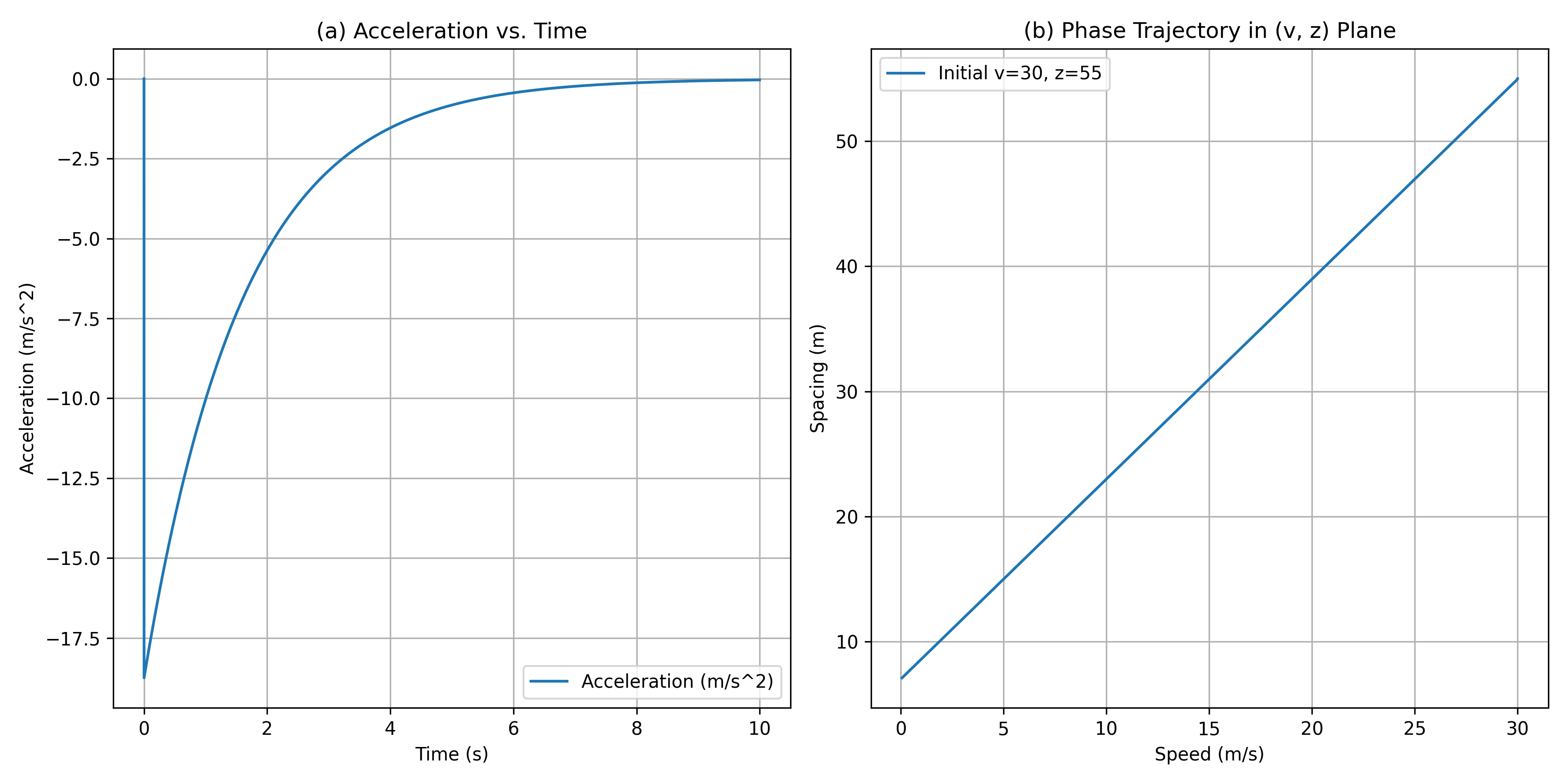}
  \caption{An example of solutions of the BA-Newell model for the stationary lead-vehicle problem}
  \label{fig:ba_newell_slvp_ec}
\efg

\subsection{Extension with bounded acceleration and deceleration}\label{subsection:BDA-Newell}

A straightforward approach to tackle the issue of unrealistically high deceleration rates present in the BA-Newell model involves directly integrating the bounded deceleration principle into the derivation process, alongside the maximum speed, minimum time gap, speed limit, and bounded acceleration principles. This is equivalent to constraining the acceleration rate in \refe{BA-Newell} from below, resulting in the following BDA-Newell model:
\bsq \label{BDA-Newell}
\begin{align}
a(t)&=\max\{-\beta,\min\{\alpha(1-\frac{v(t)}{\mu}), \frac{v_*(t)-v(t)}\epsilon\}\}, \label{BDA-Newell-a}
\end{align}
which leads to
\begin{align}
v(t+\epsilon)&=\max\{v(t)-\epsilon \beta,\min\{v(t)+\epsilon \alpha(1-\frac{v(t)}{\mu}), \mu, \frac{z(t)-\zeta}{\tau}\}\}, \label{BDA-Newell-v}\\
X(t+\epsilon)&=X(t)+ \epsilon \cdot \nonumber\\ &  \max\{v(t)-\epsilon \beta,\min\{v(t)+\epsilon \alpha(1-\frac{v(t)}{\mu}), \mu, \frac{z(t)-\zeta}{\tau}\}\}. \label{BDA-Newell-x}
\end{align}
\esq
In other words, \refe{BDA-Newell} solves the optimization problem with the objective function defined by the maximum speed principle, subject to constraints defined by the bounded acceleration, bounded deceleration, minimum time gap,  and speed limit principles. It is straightforward to show that it satisfies the fundamental diagram principle in steady states; i.e., Lemmas \ref{lemma:Newell-FD} and \ref{lemma:BA-Newell-FD} can be easily extended for the model. Consequently, the model successfully encompasses the principles of maximum speed, minimum time gap, speed limit, fundamental diagram, bounded acceleration, and bounded deceleration.

\bfg\bc \includegraphics[width=6in]{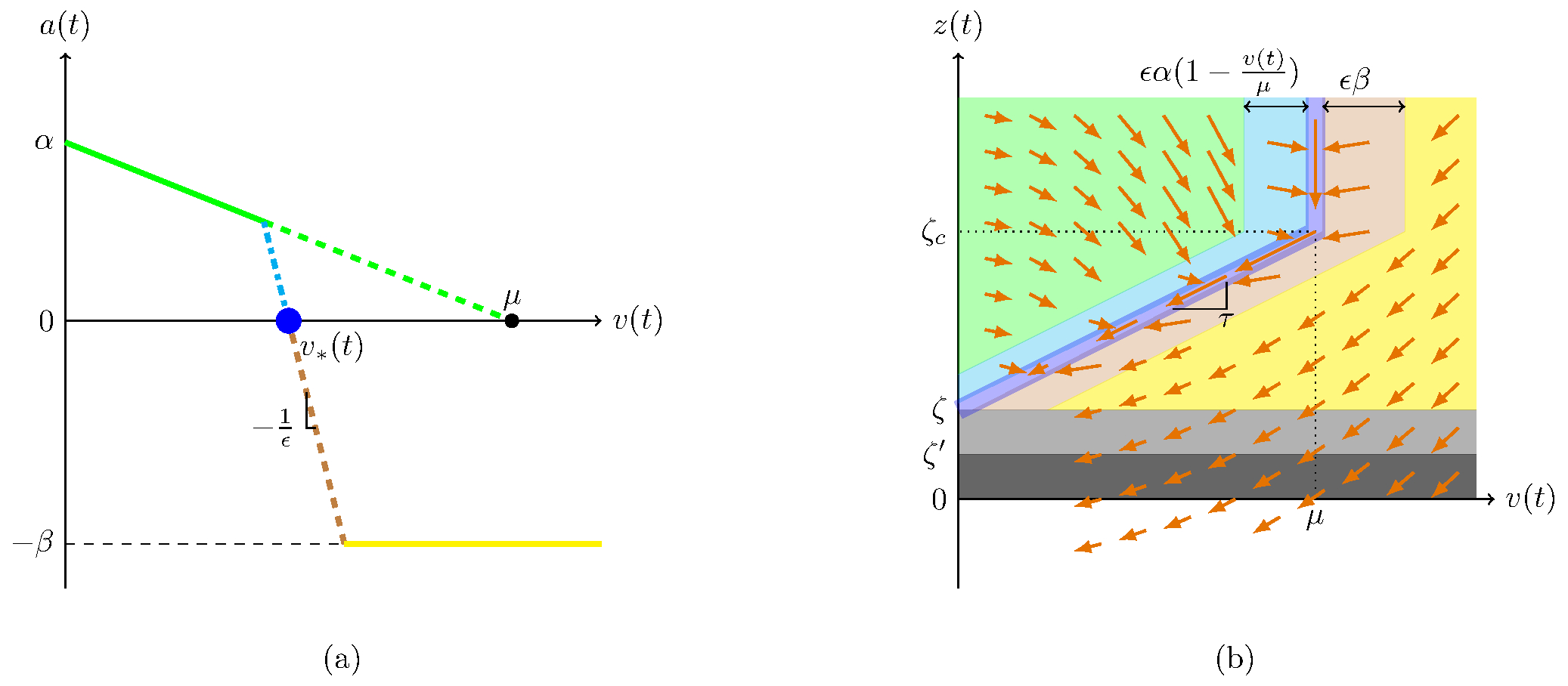} \caption{Five phases in the BDA-Newell model in \refe{BDA-Newell}}\label{BDA-Newell-5phases} \ec\efg

However, extending Lemmas \ref{lemma:Newell-CSS} and \ref{lemma:BA-Newell-CSS} to prove adherence to the forward traveling and comfort jam spacing principles becomes non-trivial, and indeed, such adherence is not guaranteed under all conditions. This is demonstrated by analyzing the phases in the model. In addition to the bounded acceleration, equilibrium cruising, equilibrium acceleration, and equilibrium deceleration phases, there is an additional bounded deceleration phase. Among these phases, the definitions of bounded acceleration, equilibrium cruising, and equilibrium acceleration phases are the same, but that of equilibrium deceleration needs to be revised: the equilibrium deceleration phase when $v_*(t)<v(t)<v_*(t)+\epsilon \beta$ and $-\beta<a(t)<0$. In addition, the bounded deceleration phase when $v(t)\geq v_*(t)+\epsilon \beta$ and $a(t)=-\beta$.  The five phases are visually represented by the green solid line (bounded acceleration), blue dot (equilibrium cruising), cyan dash-dotted line (equilibrium acceleration), brown dashed line (equilibrium deceleration), and yellow solid line (bounded deceleration) in the $(v(t),a(t))$ phase plane, as depicted in \reff{BDA-Newell-5phases}(a). In \reff{BDA-Newell-5phases}(b), the five phases are also illustrated by the correspondingly colored regions or curve in the $(v(t),z(t))$ phase plane. In the figure, the states in the grey region 
violates the comfort jam spacing principle, but not the minimum jam spacing principle; and those in the black region violates the minimum jam spacing principle.

\bfg
\centering
  \includegraphics[width=6in]{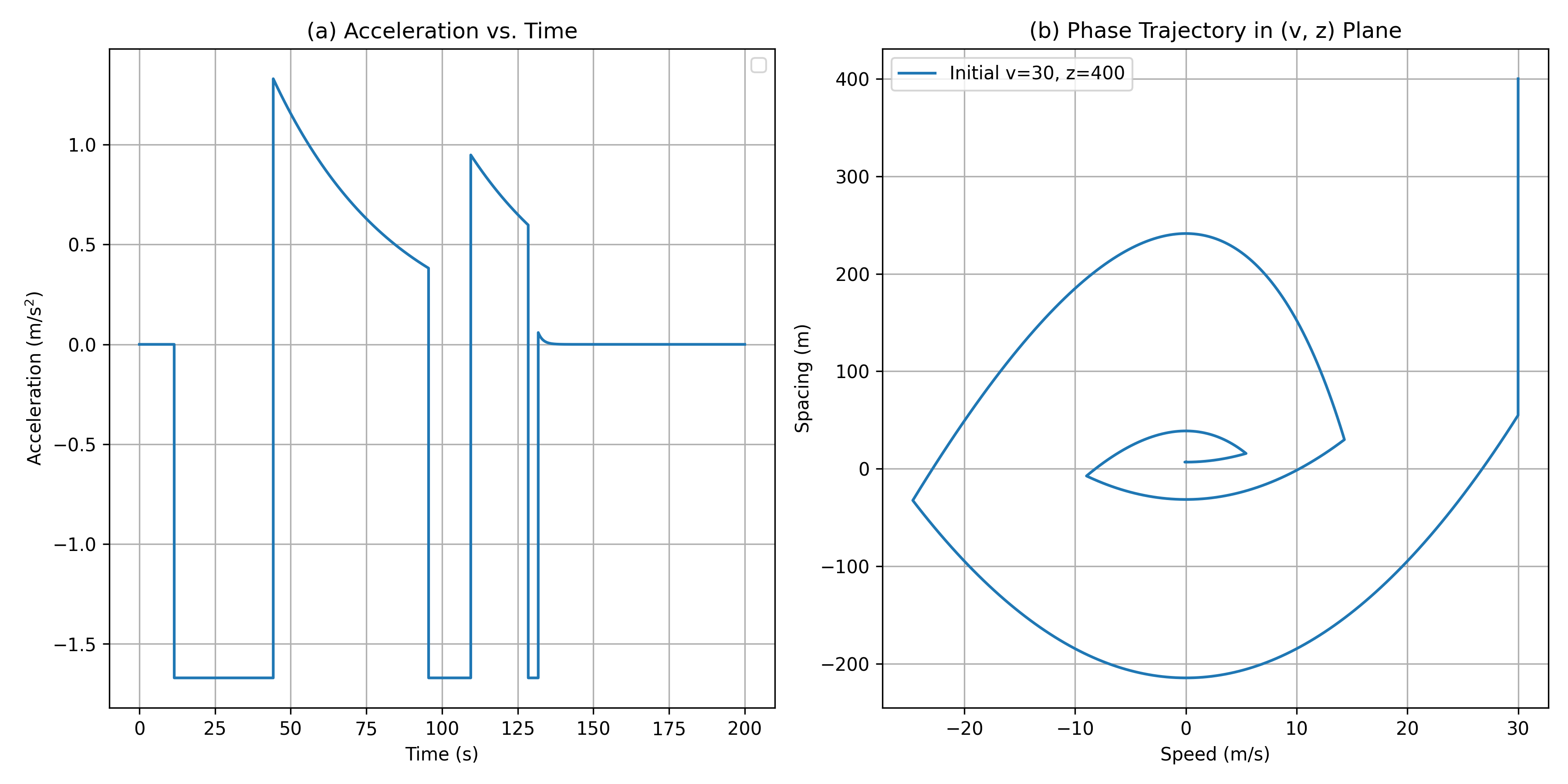}
  \caption{An example of bounded deceleration solutions of the BDA-Newell model in \refe{BDA-Newell}}
  \label{fig:bda_newell_slvp_ec}
\efg

In \reff{BDA-Newell-5phases}(b), we provide an illustration of the vector field within the $(v(t), z(t))$ phase plane for the stationary lead-vehicle problem. A noteworthy change occurs in the equilibrium deceleration phase where the acceleration $a(t)$ is bounded, resulting in solid arrows. In this new bounded deceleration phase, the vector field is characterized by $(\frac{d}{dt}v(t),\frac{d}{dt}z(t))=-(\beta,v(t))$. It is evident from the figure that, under certain initial conditions, the phase trajectory may not converge to the equilibrium cruising state. Instead, it can lead to the grey or even the black regions, indicating violations of 
the comfort jam spacing or even the minimum jam spacing principles within the model. As an example,  we set initial values at $v(0)=30$ m and $z(0)=400$ m with $\Delta t$=0.0001 s for a simulation time of 200 s. Additionally, we assign  $\beta=2$ m/s$^2$. In this case, the vehicle initially drives at the free-flow speed until it reaches $(v(11.5),z(11.5))=(30,55)$, which is the same initial state as that in \reff{fig:ba_newell_slvp_ec},  and then consistently operates within the bounded deceleration phase and comes to a halt after $18.0$ s. Consequently, the resulting spacing is calculated as $z(29.5)=-214.5$ m, based on the formula $z(t)=z(11.5)-\frac{v^2(0)}{2\beta}$. The solutions are illustrated in \reff{fig:bda_newell_slvp_ec}, from which we can see that the vehicle quickly collides into the rear of the stationary leader, and also attempts to move backward until eventually coming to a stop at the comfort jam spacing. To ensure that the model complies with 
the comfort jam spacing principle, the comfort deceleration bound $\beta$ would need to be as high as $\beta=\frac {v(0)}{2\tau}=9.375$ m/s$^2$.\footnote{It's worth noting that when $\beta=9$ m/s$^2$, the vehicle initially comes to a halt at a spacing of $\zeta'=5$ m. However, it subsequently reverses direction, traveling backward, and ultimately comes to a stop at $\zeta=7$ m.} However, this value surpasses the reasonable range for deceleration rates, as ISO 15622 sets the comfort deceleration bound at approximately 5 m/s$^2$ \citep{xiao2010comprehensive}. In such scenarios, vehicles must be adequately prepared for the potentially extensive braking distance required when the leader begins to brake or comes to a complete stop, especially if they have not maintained a safe following distance.

In addition, we would like to highlight that the BDA-Newell model prohibits initial spacings smaller than the comfort jam spacing and the minimum  jam spacing, which are depicted in the grey region in \reff{BDA-Newell-5phases}(b) and could arise for vehicles stopping at an intersection or changing lanes in highways. In such cases, vehicles could travel backwards in the BDA-Newell model. This can be illustrated by an example with $z(t)=\zeta'$, and $v(t)=0$. From \refe{BDA-Newell}, the vehicle will decelerate to a negative speed. In summary, while the BDA-Newell model successfully adheres to the bounded deceleration principle, it may inadvertently contravene the minimum jam spacing and forward traveling principles.

\section{Analysis of the Intelligent Driver Model}
In \citep{treiber2000congested}, the following acceleration equation 
\begin{align}
a(t)&=\alpha\left(1-\left(\frac{v(t)}{\mu}\right)^\delta-\left(\frac{2\sqrt{\alpha\beta}(\zeta-\zeta'+\tau v(t))+v(t)(v(t)-v_L(t))}{2\sqrt{\alpha\beta}(z(t)-\zeta')} \right)^2\right), \label{def:IDMa}
\end{align}
 coupled with \refe{def:symplectic}, leads to the Intelligent Driver Model. In the model, in addition to the six parameters defined earlier ($\zeta$, $\zeta'$, $\tau$, $\mu$, $\alpha$, and $\beta$), an additional exponent parameter, $\delta$, is used. Thus, this model uses seven parameters, but $\delta$ is generally set to be $4$.
 
While the Intelligent Driver Model has been analyzed for its braking strategy and collision avoidance properties \citep[][Section 11.3.4]{treiber2013traffic}, our detailed examination reveals important limitations not fully addressed in the existing literature. Our simulations confirm that the model generally maintains the minimum jam spacing principle through its self-regulating braking strategy. However, in this section, we highlight two critical limitations: (1) the violation of the forward traveling principle through the emergence of backward movement during final approach, which we demonstrate through both linearization analysis and numerical simulation, and (2) the initiation of braking at distances far exceeding what would be considered human-like according to the safe stopping distance principle. These limitations are particularly relevant for applications in mixed-autonomy environments and infrastructure designed based on human driving patterns. Our analysis applies to both the original model in \citep{treiber2000congested} and the slightly modified version in \citep[][Section 11.3]{treiber2013traffic}.

\subsection{Analytical results for the stationary lead-vehicle problem}

As pointed out in \citep{albeaik2022limitations}, if the initial spacing $z(t)$ is larger than the minimum jam spacing, $\zeta'$, but smaller than the comfort jam spacing, $\zeta$, then the follower needs to travel backwards. This can be shown for the extreme case when both the leader and the follower are initially stationary at $t$; i.e., when $v(t)=v_L(t)=0$: if $z(t)-\zeta'< \zeta -\zeta'$, then $a(t)=\alpha \left(1- \left(\frac{\zeta-\zeta'}{z(t)-\zeta'}\right)^2\right)<0$, and the follower travels backwards in the Intelligent Driver Model.

In the following, we demonstrate that the follower travels backward regardless of the initial conditions for the stationary lead-vehicle problem, where $v_L(t)=0$. In this case, the model can be re-written as
\bsq
\begin{align} \label{IDM-SLVP}
\frac{d}{dt} v(t)&=\alpha\left(1-\left(\frac{v(t)}{\mu}\right)^4-\left(\frac{2 \sqrt{\alpha\beta} (\zeta-\zeta'+\tau v(t))+v^2(t)}{2 \sqrt{\alpha\beta} (\zeta-\zeta'+\tilde z(t)) } \right)^2\right),\\
\frac{d}{dt} \tilde z(t)&=-v(t),
\end{align}
\esq
where the clearance $\tilde z=z(t)-\zeta$.
The system \refe{IDM-SLVP} has an equilibrium with $(v,\tilde z)=(0,0)$, when the follower stops at a spacing of $\zeta$. 
We can linearize the system as 
\bsq
\begin{align} \label{IDM-SLVP-linear}
\frac{d}{dt} v(t)&\approx - 2 \tau \frac{\alpha}{\zeta-\zeta'} v(t)+ 2 \frac{\alpha}{\zeta-\zeta'} \tilde z(t) ,\\
\frac{d}{dt} \tilde z(t)&=-v(t),
\end{align}
\esq
whose Jacobian matrix is 
\begin{align*}
J &=
\begin{bmatrix}
- 2 \tau \frac{\alpha}{\zeta-\zeta'} & 2  \frac{\alpha}{\zeta-\zeta'}\\
-1 & 0 
\end{bmatrix}.
\end{align*}
We can find its eigenvalues as
\begin{align*}
\lambda_{1,2} &= \frac{\alpha}{\zeta-\zeta'}\left(- \tau  \pm \sqrt{\tau ^2  -  2 \frac{\zeta-\zeta'}{\alpha}}\right).
\end{align*}
With the values in Table 1 of \citep{treiber2000congested}: $\zeta=7$ m, $\zeta'=5$ m, $\tau=1.6$ s, and $\alpha=0.73$ m/s$^2$, we have $\lambda_{1,2}=  -0.584 \pm 0.624i$. According to \citep[][Page 134]{strogatz1994nonlinear}, the approximate linear system in \refe{IDM-SLVP-linear} converges to the equilibrium point with a stable spiral. During the process, the speed $v(t)$ inevitably becomes negative, regardless of the initial conditions.

Still for the stationary lead-vehicle problem, if the follower's initial speed is the free-flow speed, $v(t)=\mu$, from \refe{def:IDMa}, we can see that $a(t)<0$, regardless of $z(t)$. That is, however far away the vehicle is from a traffic light (as a stationary leader), it immediately decelerates to stop if the light is red. Therefore, the braking distance can be arbitrarily long in the Intelligent Driver Model. Thus, it violates the safe stopping distance principle in \refe{SSD-principle}.

\subsection{A numerical example}

In the following, we replicate the example in Figure 2 of \citep{treiber2000congested} to highlight the back traveling and extremely long stopping distance issues identified in the preceding subsection. Our simulation uses the same parameters as specified in \citet{treiber2000congested} and ISO 15622 \citep{hiraoka2005modeling}: $\Delta t=0.001$ s, comfort jam spacing $\zeta=7$ m, minimum jam spacing $\zeta'=5$ m, time headway $\tau=1.6$ s, desired speed $\mu=120$ km/h, comfort acceleration bound $\alpha=0.73$ m/s$^2$, and comfort deceleration bound $\beta=1.67$ m/s$^2$.

\bfg
\centering
  \includegraphics[width=6in]{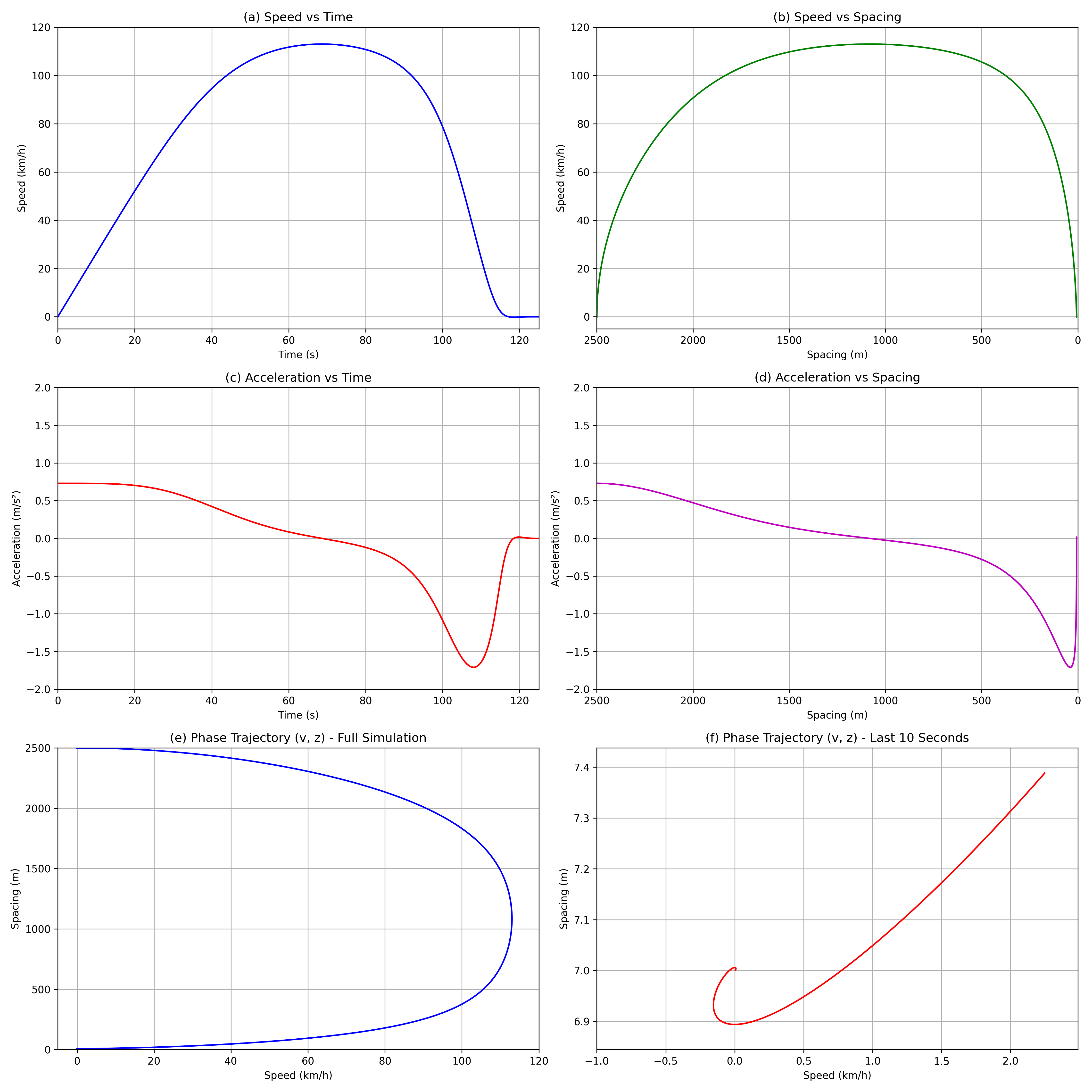}
  \caption{Replication of Figure 2 in \citep{treiber2000congested}}
  \label{fig:treiber2000congested_example2_125s}
\efg

In \reff{fig:treiber2000congested_example2_125s}, we present a comprehensive replication and extension of the scenario originally shown in Figure 2 of \citep{treiber2000congested}. While the original publication included only the speed-time and acceleration-time profiles (panels (a) and (c) in our figure), we have expanded the analysis to include speed-spacing relationships, acceleration-spacing relationships, and phase trajectories to provide deeper insights into the IDM's behavior. The six panels show: (a) speed versus time, (b) speed versus spacing, (c) acceleration versus time, (d) acceleration versus spacing, (e) phase trajectory in the speed-spacing plane for the entire simulation, and (f) phase trajectory for the last 10 seconds.

Several notable observations emerge from this extended analysis: 
\bi
\item First, panels (b) and (d) reveal one of the model's key limitations—the vehicle begins decelerating at distances exceeding 1000 meters from the stationary leader. This far exceeds the safe stopping distance calculated from \refe{SSD-principle} with $\tau' = 1$ s, $\beta = 1.67$ m/s$^2$, and a maximum speed of $v = 120$ km/h, which is about 366 m.
Thus, the IDM initiates braking at approximately 1000 meters, which is about 2.7 times greater than the safe stopping distance. Even when accounting for a gradual deceleration profile, this discrepancy indicates an overly conservative braking behavior not aligned with typical human driving patterns.
\item Second, panel (f) provides critical insight into the model's approach behavior. The trajectory in the speed-spacing plane converges to the equilibrium point $(0,7)$—representing a stopped vehicle maintaining the comfort spacing of 7 meters—but it does so through a stable spiral. This spiral pattern reveals that the vehicle's speed oscillates around zero during the final approach, temporarily dropping below zero. This negative speed indicates an unrealistic backward driving behavior of the vehicle, highlighting a significant limitation in the model's behavior during the final stopping phase.
\ei

These observations confirm the theoretical concerns regarding excessive braking distances and unrealistic back-traveling phenomena discussed in the preceding subsection. The excessive braking distance of more than 1000 meters could lead to safety concerns at signalized intersections where the change and clearance intervals are designed for normal human-driven vehicles, while the backward traveling during final approach represents physically implausible vehicle behavior.

\section{Analysis of the Gipps model}
In \citep{gipps1981bcf}, a car-following model was derived based on the safe stopping distance principle in \refe{SSD-principle}, such that
  \begin{align}
(\tau'-\tau_1) v(t)+ \tau_1 v(t+\epsilon)  +\frac{v^2(t+\epsilon)}{2\beta} - \frac{v_L^2(t)}{2\beta} \leq z(t)-\zeta, \label{Gipps-SSD}
  \end{align}
which leads to
\begin{align}
v(t+\epsilon) &\leq -\beta\tau_1 + \sqrt{\beta^2\tau_1^2 + 2\beta(z(t)-\zeta) + 2\beta(\tau_1 - \tau')v(t) +  v_L^2(t)}.
\end{align}
Coupled with the bounded acceleration model, we have
\begin{align}
v(t+\epsilon) &= \min\{ v(t)+ 2.5 \alpha' \epsilon (1-v(t)/\mu) \sqrt {0.025+v(t)/\mu}, \nonumber\\
&-\beta\tau_1 + \sqrt{\beta^2\tau_1^2 + 2\beta(z(t)-\zeta) + 2\beta(\tau_1 - \tau')v(t) +  v_L^2(t)}\}.
\end{align}
In the original derivation, $\tau'=1$ s, $\tau_1=2/3$ s, and $\epsilon=2/3$ s. For Krauss' model and other variations of the Gipps model, refer to \citep{Krauss1998microscopic}.

Here we consider a simplified version of \refe{Gipps-SSD} with $\tau_1=\tau'$:
  \begin{align}
\tau' v(t+\epsilon)  +\frac{v^2(t+\epsilon)}{2\beta} - \frac{v_L^2(t)}{2\beta} \leq z(t)-\zeta, \label{simplified-Gipps-SSD}
  \end{align}
which leads to
\begin{align}
v(t+\epsilon) &\leq -\beta\tau' + \sqrt{\beta^2\tau'^2 + 2\beta(z(t)-\zeta)  +  v_L^2(t)}.
\end{align}
Coupled with the TWOPAS bounded acceleration model, we have the following simplified Gipps model
\begin{align}
v(t+\epsilon) &= \min\{ v(t)+ \epsilon \alpha  (1-\frac{v(t)}{\mu}), -\beta\tau' + \sqrt{\beta^2\tau'^2 + 2\beta(z(t)-\zeta)  +  v_L^2(t)}\}. \label{simplified-Gipps-model}
\end{align}
This model has five parameters: $\zeta$, $\tau'$, $\mu$, $\alpha$, and $\beta$.
Here $\epsilon$ can be any time-step size $\Delta t$, thus making this a continuous model. Note that the simplified Gipps model  in \citep[][Section 11.2]{treiber2013traffic} is a special discrete case with $\epsilon=\tau'$.

\subsection{Analytical solutions and properties}
In \refe{simplified-Gipps-model}, the term under the square root is non-negative when $z(t)\geq \zeta$. However, for spacings $z(t)$ such that $\zeta' \le z(t) < \zeta$, the term under the square root can become negative if $v_L(t)$ is sufficiently small, rendering the model ill-defined in those conditions. Thus, the model cannot be extended for lane-changing traffic or arterial road traffic, where cars can stop or even travel at a smaller spacing than the comfort jam spacing.

In the steady state where $v(t+\epsilon)=v(t)=v_L(t)=v$, we have the following speed-density relation \citep[][Section 11.2]{treiber2013traffic}:
\begin{align}
v&=\min\{\mu, \frac1 {\tau'} (\frac 1k-\frac 1\kappa)\}, \label{Gipps-triangular-vk}
\end{align}
which leads to the triangular fundamental diagram. Note that, different from that in \refe{triangular-vk}, the shock wave speed is $\frac {1}{\tau' \kappa}$, not $\frac {1}{\tau \kappa}$ as that in Newell's simplified car-following model.

The simplified Gipps model, \refe{simplified-Gipps-model}, has two phases. In the first phase, $\dot v(t)=\alpha  (1-\frac{v(t)}{\mu})$, in which the vehicle accelerates when $v(t)<\mu$, decelerates when $v(t)>\mu$, and cruises at $v(t)=\mu$. In the second phase, $v(t)=\lim_{\epsilon\to 0} -\beta\tau' + \sqrt{\beta^2\tau'^2 + 2\beta(z(t)-\zeta)  +  v_L^2(t)}$. Since $\dot z(t)=v_L(t)-v(t)$, we have 
\begin{align*}
\dot z(t)&=v_L(t)+\beta\tau' - \sqrt{\beta^2\tau'^2 + 2\beta(z(t)-\zeta)  +  v_L^2(t)}\\&=\beta\tau'- \frac{\beta^2\tau'^2 + 2\beta (z(t)-\zeta)}{v_L(t)+ \sqrt{\beta^2\tau'^2 + 2\beta(z(t)-\zeta)  +  v_L^2(t)}}.
\end{align*}
Thus, when $z(t)\geq \zeta$, $\dot z(t)$ attains the minimum value for $v_L(t)=0$, and the spacing is reduced at the fastest rate for a stationary leader. Thus, we have 
\begin{align*}
\dot z(t)&\geq \beta\tau'-  \sqrt{\beta^2\tau'^2 + 2\beta(z(t)-\zeta) },
\end{align*}
and the lower bound of $z(t)$ satisfies the following first-order nonlinear differential equation:
\begin{align}
\frac{dz}{dt} = \beta\tau' - \sqrt{\beta^2\tau'^2 + 2\beta(z-\zeta)}.
\end{align}
We solve this differential equation to obtain $t$ as a function of $z$:
\begin{align}
t(z) = -\frac{\sqrt{\beta^2\tau'^2 + 2\beta(z-\zeta)}}{\beta} - \tau' \ln\left|\frac{\beta\tau' - \sqrt{\beta^2\tau'^2 + 2\beta(z-\zeta)}}{v(0)}\right| + \frac{v(0) + \beta\tau'}{\beta},
\end{align}
where the initial speed $v(0) \in(0, \mu]$, and the initial spacing from the stationary leader is given by:
\begin{align}
z(0) = \zeta + v(0)\tau'+\frac{v^2(0) }{2\beta}.
\end{align}
Thus, the safe stopping distance from $v(0)$ is consistent with that in \refe{SSD-principle}; and the spacing approaches the comfort jam spacing when $t\to \infty$.
In addition, the speed as a function of position is:
\begin{align}
v(z) = -\beta\tau' + \sqrt{\beta^2\tau'^2 + 2\beta(z-\zeta)},
\end{align}
and the acceleration as a function of position is:
\begin{align}
a(z) = -\beta + \beta \frac{\beta\tau'}{\sqrt{\beta^2\tau'^2 + 2\beta(z-\zeta)}}.
\end{align}
We can see that (i) the speed is always non-negative and approaches $0$ when $z\to \zeta$; and (ii) $a(z)\geq -\beta$ and increases to $0$ when $z\to \zeta$; i.e., the deceleration is bounded by $\beta$.

From the above analysis, we can see that the simplified Gipps model satisfies all of the principles defined in Section 2, except that: (i) the model may not be well-defined for a spacing between the comfort and minimum jam spacings; (ii) the fundamental diagram and the corresponding shock wave speed are inconsistent with human-driven vehicles. If setting $\tau'$ to be the same as $\tau$, one can resolve the second limitation, but then the safe stopping distance is no longer consistent with that in \refe{SSD-principle}. In summary, the model is simple and well-behaved, but lacks sufficient flexibility.

\subsection{A numerical example}

In this subsection, we replicate the example in Figure 2 of \citep{treiber2000congested} with the simplified Gipps model, \refe{simplified-Gipps-model}. Our simulation uses the same parameters: $\Delta t=0.001$ s, comfort jam spacing $\zeta=7$ m, reaction time $\tau'=1$ s, desired speed $\mu=120$ km/h, comfort acceleration bound $\alpha=0.73$ m/s$^2$, and comfort deceleration bound $\beta=1.67$ m/s$^2$. 

\bfg
\centering
  \includegraphics[width=6in]{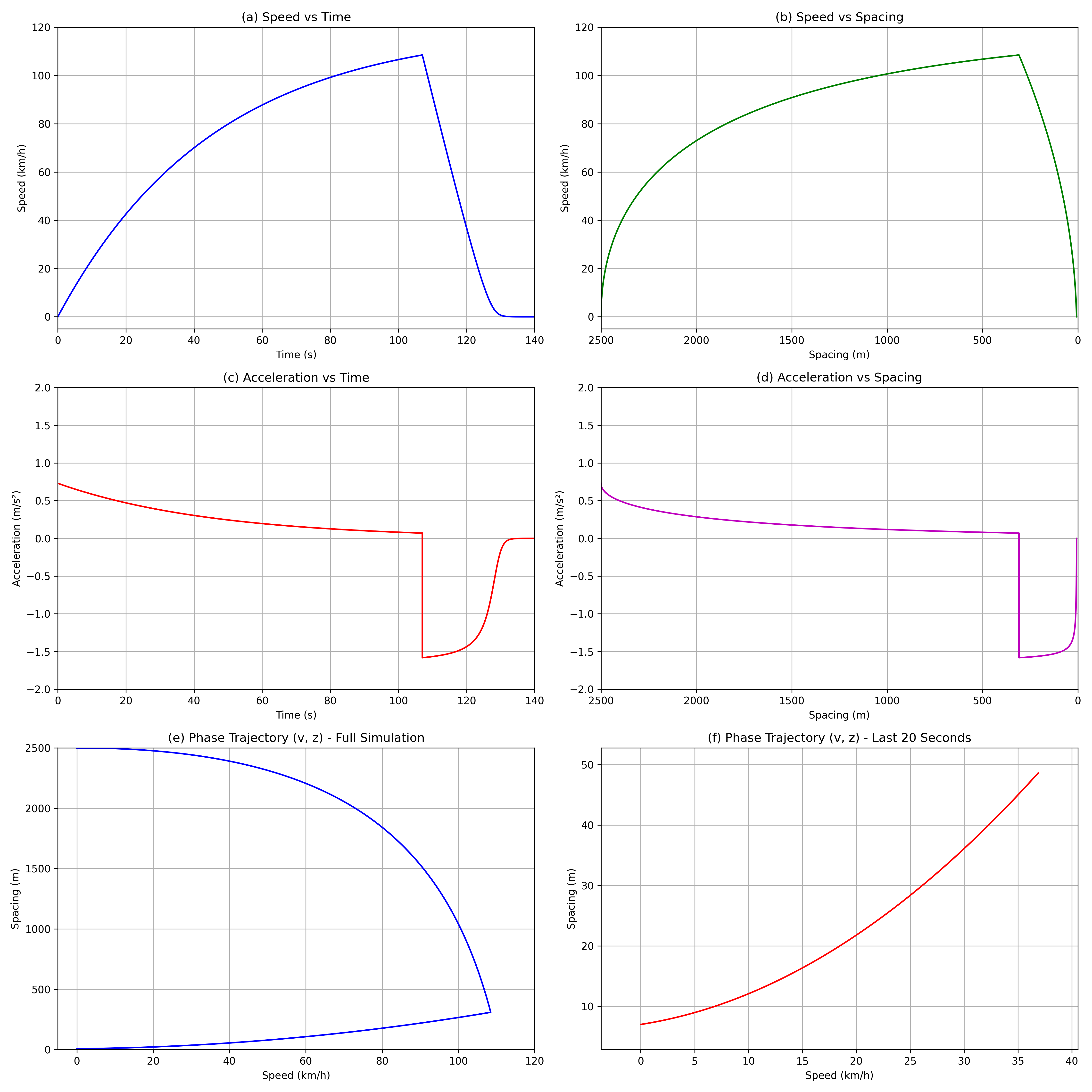}
  \caption{Replication of Figure 2 in \citep{treiber2000congested} with the simplified Gipps model}
  \label{fig:simplified_Gipps_140s}
\efg

In \reff{fig:simplified_Gipps_140s}, the six panels show: (a) speed versus time, (b) speed versus spacing, (c) acceleration versus time, (d) acceleration versus spacing, (e) phase trajectory in the speed-spacing plane for the entire simulation, and (f) phase trajectory for the last 20 seconds.
From the figures we can see that: 
\bi
\item First, panel (f) confirms that the trajectory in the speed-spacing plane converges to the equilibrium point $(0,7)$, and the speed is always non-negative.
\item Second, panels (a) and (b) show that the vehicle accelerates to the maximum speed of about 108 km/h with the first phase of the model, and then enters the second phase to brake to stop. The stopping distance is about 301 m, consistent with that in \refe{SSD-principle}.
\item Third, a closer look at panels (c) and (d) reveals that, when the vehicle switches from the first phase to the second, there is a sharp change in the acceleration rate, from a positive value to a minimum value of about -1.6 m/s$^2$. This suggests that the simplified Gipps model has no bound on the jerk, which is another limitation.
\ei

These results verify the analytical results in the preceding subsection, and further reveal another limitation related to infinite jerks.

\section{Conclusion}
In this article, we introduced and applied a multi-phase dynamical systems analysis framework to systematically evaluate car-following models against fundamental principles of safe and human-like driving. We began by formulating these principles, encompassing zeroth-order comfort and minimum jam spacings, first-order speed and time gap considerations, and second-order acceleration/deceleration bounds and braking profiles. Using this framework, we derived Newell's simplified model and subsequently analyzed its extensions, along with the Intelligent Driver Model and the Gipps model, focusing on the stationary lead-vehicle problem.

Our multi-phase analysis in the speed-spacing plane, a key methodological contribution of this work, revealed critical insights into the behavior of these standard models. This systematic phase-plane analysis enabled a rigorous examination that uncovered previously under-appreciated limitations. For instance, Newell's model, while parsimonious, requires bounded acceleration for realistic behavior; however, naive extensions to include bounded deceleration can lead to violations of minimum jam spacing and forward travel principles. Similarly, our analysis highlighted that the Intelligent Driver Model can exhibit unrealistic backward travel near equilibrium and excessively long braking distances. The Gipps model, while generally adhering to safety principles under certain conditions, becomes ill-defined for some spacings and has inconsistencies in its fundamental diagram parameters if aligned with safe stopping distances. These analytical insights, validated by numerical simulations, underscore the limitations of existing models in consistently satisfying all defined principles for safe and human-like driving.

In Part 2 of this study \citep{jin2025WA20-02_Part2}, we directly address these limitations by developing a novel multi-phase projection-based car-following model. Building upon the analytical framework established here, Part 2 introduces the concept of projected braking to anticipate leader vehicle dynamics, providing a parsimonious solution that maintains bounded acceleration/deceleration while ensuring collision-free operation under realistic initial conditions.

Both parts of this study highlight the utility of multi-phase dynamical systems analysis for a deeper understanding of car-following model behaviors and their adherence to desired driving principles. The identified limitations in standard models motivate the need for new approaches that can rigorously integrate safety, human-like characteristics, and operational efficiency. For future research, the overarching goal is to develop car-following models that not only conform to all established principles of safe and human-like driving but are also parsimonious in their parameterization and flexible enough to capture diverse driving behaviors and adapt to varied scenarios. This involves pursuing models that can be rigorously proven safe, yet are computationally efficient and require minimal parameters for calibration, making them suitable for real-world learning and deployment in automated vehicles. Future work should focus on creating new models or enhancing existing ones to achieve this crucial balance. Extending the multi-phase analysis to more complex traffic scenarios, such as interactions at signalized intersections, merging, and operations in mixed traffic with varying levels of vehicle automation, remains a vital direction. Furthermore, empirical validation and parameter calibration using diverse real-world trajectory datasets will be indispensable for developing models that are not only theoretically sound but also practically applicable for both simulating human driving accurately and guiding the design of intelligent and safe autonomous vehicle control systems.

\section*{Acknowledgments}

The author expresses sincere gratitude to the UC ITS Statewide Transportation Research Program (STRP) for their generous financial support. Special thanks are extended to Dr. Ximeng Fan for valuable discussions on the model, data, and related literature. The author also thanks Dr. Zuduo Zheng of the University of Queensland and the anonymous reviewers of Transportation Research Part B for their constructive feedback on an earlier version of this manuscript. The author appreciates the assistance of AI language models during the iterative process of revising text and exploring concepts, which contributed to the efficiency of this work. Any presented results, perspectives, or errors are the sole responsibility of the author.

\section*{Declaration of generative AI and AI-assisted technologies in the writing process}
During the preparation of this work the author used Claude, ChatGPT, and Google Gemini in order to improve language clarity, enhance readability, and refine phrasing in specific sections. After using these tools, the author reviewed and edited all content thoroughly and takes full responsibility for the content of the publication.

\end{document}